\documentclass[onecolumn,secnumarabic,amssymb, nobibnotes, aps, prd]{revtex4}
\usepackage{amsmath} \usepackage{amssymb}
\usepackage{graphicx} 
\usepackage{epstopdf}
\usepackage{amsmath}
\usepackage{amsmath,amssymb,amsthm,amsfonts,mathrsfs,bm,verbatim}
\usepackage{graphicx,subfigure}

\newcommand{\bea}{\begin{eqnarray}}
\newcommand{\eea}{\end{eqnarray}}

\begin{document}

\title{Analysis on superradiant instability of charged dilaton Black Hole }%
\author{Wen-Xiang Chen$^{a}$}
\affiliation{Department of Astronomy, School of Physics and Materials Science, GuangZhou University, Guangzhou 510006, China}
\author{Yao-Guang Zheng}
\email{hesoyam12456@163.com}
\affiliation{Department of Astronomy, School of Physics and Materials Science, GuangZhou University, Guangzhou 510006, China}

\begin{abstract}
In this post, a new variable y($\mu=y\omega$) is added to extend the results of the above post. We derive a very compact resonance formulation. When y is a certain limit, the effective potential of the equation exists at the pole, and the potential well appears outside the event horizon, when  ${e}^{-4\alpha \Phi /(n-1)}\sqrt{\alpha} Q \ll r_{+}\ll 1 / (\frac{4}{n-1}y) $, so charged dilator black hole is superradiatively unstable at this time.
\end{abstract}

\maketitle

\section{Introduction}
The black hole no hair theorem was first proposed by Wheeler in 1971 \cite{1}.In the 1970s, the development of black hole thermodynamics applied the fundamental laws of thermodynamics to the theory of black holes in the field of general relativity, strongly suggesting that the general relativity, thermodynamics and quantum theory.
The stability of black holes is an important topic in black hole physics.Regge and Wheeler\cite{2} demonstrated that spherically symmetric Schwarzschild black holes are stable under perturbations. The stability of rotating black holes is further complicated by the large effect of superradiation. Superradiation effects occur in both classical and quantum scattering processes. When a boson wave hits a rotating black hole, it is possible that the rotating black hole is as stable as a Schwarzschild black hole if certain conditions are met.\cite{1,2,3,4,5,6,7,8,9,10,11}
\begin{equation}
    \omega< m\varOmega_H+q\varPhi_H,  \varOmega_H=\frac{a}{r_+^2+a^2}
\end{equation}
where $q$ and $m $ are the charge and azimuthal quantum number of the incoming wave, $\omega$ denotes the wave frequency, $\varOmega_H$ is the angular velocity of black hole horizon and $\varPhi_H$ is the electromagnetic potential of the black hole horizon. If the frequency range of the wave lies in the superradiant condition, the wave reflected by the event horizon will be amplified, which means that the wave extracts rotational energy from the rotating black hole during the scattering of the incident wave. According to the black hole bomb mechanism proposed by Press and Teukolsky (\cite{1,2,3,4,5,6,7,8,9}), if a mirror is placed between the event dome and the outer space of the black hole, the amplified waves will be reflected back and forth between the mirror and the black hole and grow exponentially, which leads to the superradiant instability of the black hole.

 Dilaton gravity \cite{12,13,14} has received a lot of attention from the community because of its great physical significance. First, it is the low-energy limit of string theory because in the low-energy limit one can recover Einstein gravity with non-minimally coupled dilaton fields and other fields. Secondly, its action consists of one or more Liouville-type potentials. These potentials can be generated by spacetime supersymmetry breaking in ten dimensions. In addition, it has been reported that dilatant fields may affect the stochastic structure and thermodynamic properties of black holes. Due to the physical importance of this point, various black hole solutions have been proposed and their thermodynamic properties have been discussed . Interestingly, it was also proposed  where it is proposed that the cosmological constant is found to be ory.

In this post, a new variable y($\mu=y\omega$) is added to extend the results of the above post. We derive a very compact resonance formulation. When y is a certain limit, the effective potential of the equation exists at the pole, and the potential well appears outside the event horizon, when  ${e}^{-4\alpha \Phi /(n-1)}\sqrt{\alpha} Q \ll r_{+}\ll 1 / (\frac{4}{n-1}y) $, so charged dilator black hole is superradiatively unstable at this time.

\section{The system of  charged dilaton black hole}

The Einstein-Maxwell-Dilaton action in $(n+1)$-dimensional spacetime reads \cite{8}
\begin{equation}
\begin{aligned}
S=& \frac{1}{16 \pi} \int \mathrm{d}^{n+1} x \sqrt{-g}\left(\mathcal{R}-\frac{4}{n-1}(\nabla \Phi)^{2}-V(\Phi)\right.\\
&\left.-\mathrm{e}^{-4 \alpha \Phi /(n-1)} F_{\mu \nu} F^{\mu \nu}\right)
\end{aligned}
\end{equation}
where ${R}, \Phi$, and $F_{mu\nu}$ denote the Ricci scalar curvature, the expansion field, and the electromagnetic field tensor, respectively. The parameter $alpha$ describes the strength of the coupling between the electromagnetic and sparse fields.

In order to obtain a solution, we must adopt a particular form of the dilution field $V(\Phi)$; in this case, $V(\Phi)=2 \Lambda_{0} \mathrm{e}^{2 \zeta_{0} \Phi}+2 \Lambda \mathrm{e}^{2 \zeta \Phi}$. Note that if the power-law Maxwell field is considered, an additional term should be added to the above potential.

We can think of the general form of the metric system as\cite{16}.
\begin{equation}
\mathrm{d} s^{2} = -g(r) \mathrm{d} t^{2}+\frac{1}{g(r)} \mathrm{d} r^{2}+r^{2} R^{2}(r) h_{i j} \mathrm{~d} x^{i} \mathrm{~d} x^{j},
\end{equation}
where $h_{i j} \mathrm{~d} x^{i} \mathrm{~d} x^{j}$ corresponds to an $(n-1)$-dimensional hypersurface with constant scalar curvature $(n-1)(n-2) k$, where $k$ can be taken as $-1,0,1$.

Using the theorem $R={e}^{2 \alpha \Phi /(n-1)}$, the solution can be obtained.
\begin{equation}
\begin{aligned}
&\begin{aligned}
f(r)=&-\frac{k(n-2)\left(1+\alpha^{2}\right)^{2} b^{-2 \gamma} r^{2 \gamma}}{\left(\alpha^{2}-1\right)\left(\alpha^{2}+n-2\right)}-\frac{m}{r^{(n-1)(1-\gamma)-1}} \\
&+\frac{2 \Lambda b^{2 \gamma}\left(1+\alpha^{2}\right)^{2} r^{2(1-\gamma)}}{(n-1)\left(\alpha^{2}-n\right)} \\
&+\frac{2 q^{2}\left(1+\alpha^{2}\right)^{2} b^{-2(n-2) \gamma} r^{2(n-2)(\gamma-1)}}{(n-1)\left(n+\alpha^{2}-2\right)}
\end{aligned}\\
&\Phi(r)=\frac{(n-1) \alpha}{2\left(\alpha^{2}+1\right)} \ln \left(\frac{b}{r}\right)
\end{aligned}
\end{equation}
where $b$ is an arbitrary constant and $\gamma$ is related to $\alpha$ by $\gamma=\alpha^{2} /\left(\alpha^{2}+1\right)$ and
\begin{equation}
\Lambda_{0}=\frac{k(n-1)(n-2) \alpha^{2}}{2 b^{2}\left(\alpha^{2}-1\right)}, \quad \zeta_{0}=\frac{2}{\alpha(n-1)}, \quad \zeta=\frac{2 \alpha}{n-1}
\end{equation}

 The following covariant Klein-Gordon equation
\begin{equation}
( \nabla ^{\nu}-iqA^{\nu})( \nabla _{\nu}-iqA_{\nu}) \Phi =\mu ^2\Phi,
\end{equation}
where $\nabla ^{\nu}$ represents the covariant derivative under the Kerr-Newman background. We adopt the method of separation of variables to solve the above equation, and it is decomposed as
\begin{equation}
\Phi ( t,r,\theta ,\phi) =\sum_{lm}R_{lm}( r ) S_{lm}( \theta) e^{im\phi}e^{-i\omega t}.
\end{equation}
where $R_{lm}$ are the equations which satisfy the radial equation of motion. The angular function $S_{lm}$ denote the scalar spheroidal harmonics which satisfy the angular part of the equation of motion. $l(=0,1,2,...)$ and $m$ are integers, $-l\leq m\leq l$ and $\omega$ denote the angular frequency of the scalar perturbation.

We get the following radial wave equation(another\ radial\ function\ $ \psi =\text{rR}$)
\begin{equation}
\frac{{{\text{d}}^{2}}\psi }{\text{dr}_{*}^{\text{2}}}+V\psi =0,
\end{equation},
V can change to \cite{15,17}
\begin{equation}
V(r)=g(r)\left[\mu^{2}+\frac{l(l+1)}{r^{2}}+\frac{2 M}{r^{3}}-{e}^{-4\alpha \Phi /(n-1)}\frac{\alpha Q^{2}}{r^{4}}\right]
\end{equation}

\section{The limit $y$ of the incident particle under the superradiance of charged dilaton black holes}
In this section, we will use analytical techniques to solve the Schrodinger-like ordinary differential equation, which determines the spatial behavior of non-minimally coupled large-scale scalar field configurations in the state $r>R$.\cite{18,19}

In particular, we will explicitly show that differential equations can be processed analytically in two radial regions ${e}^{-4\alpha \Phi /(n-1)}\sqrt{\alpha} Q \ll r_{+}\ll 1 / (\frac{4}{n-1}y)$.

We shall first analyze the Schrödinger-like ordinary differential equation of the non-minimally coupled massive scalar field in the radial region
\begin{equation}
r_{+}\ll 1 / (\frac{4}{n-1}y).
\end{equation}
The general mathematical solution of the differential equation can be expressed in terms of the Bessel functions of the first and second kinds
\begin{equation}
\psi(r)=A_{1} r^{\frac{1}{2}} J_{l+\frac{1}{2}}\left(\frac{{e}^{-4\alpha \Phi /(n-1)}\sqrt{\alpha} Q}{r}\right)+A_{2} r^{\frac{1}{2}} Y_{l+\frac{1}{2}}\left(\frac{\sqrt{{e}^{-4\alpha \Phi /(n-1)}\sqrt{\alpha} Q}}{r}\right),
\end{equation}
where the coefficients $\left\{A_{1}, A_{2}\right\}$ are normalization constants which, using a matching procedure, will be determined below.
And
\begin{equation}
J_{v}(z)=\frac{(z / 2)^{v}}{\Gamma(\nu+1)} \cdot\left[1+O\left(z^{2}\right)\right]
\end{equation}
and
\begin{equation}
Y_{v}(z)=-\frac{\Gamma(v)}{\pi(z / 2)^{v}} \cdot\left[1+O\left(z^{2}\right)\right]
\end{equation}
We get the mathematical expression
\begin{equation}
\psi(r)=A_{1} \frac{({e}^{-4\alpha \Phi /(n-1)}\sqrt{\alpha} Q / 2)^{l+\frac{1}{2}}}{\Gamma\left(l+\frac{3}{2}\right)} \cdot r^{-l}-A_{2} \frac{\Gamma\left(l+\frac{1}{2}\right)}{\pi(\sqrt{{e}^{-4\alpha \Phi /(n-1)}\sqrt{\alpha} Q} / 2)^{l+\frac{1}{2}}} \cdot r^{l+1}
\end{equation}
for the radial eigenfunction of the non-minimally coupled massive scalar field configurations in the intermediate region
\begin{equation}
{e}^{-4\alpha \Phi /(n-1)}\sqrt{\alpha} Q \ll r_{+}\ll 1 / (\frac{4}{n-1}y)
\end{equation}
We shall next analyze the Schrödinger-like ordinary differential equation in the radial region
\begin{equation}
{e}^{-4\alpha \Phi /(n-1)}\sqrt{\alpha} Q \ll r_{+}
\end{equation}

In terms of the Bessel functions of the first and second kinds, the general mathematical solution  can be expressed in the compact form:
\begin{equation}
\psi(r)=B_{1} r^{\frac{1}{2}} J_{l+\frac{1}{2}}(i \mu r)+B_{2} r^{\frac{1}{2}} Y_{l+\frac{1}{2}}(i \mu r),
\end{equation}
where the coefficients $\left\{B_{1}, B_{2}\right\}$ are normalization constants to be determined below.
Using the small-argument ( $\mu r \ll 1$ ) functional behaviors of the modified Bessel functions, one obtains the mathematical expression
\begin{equation}
\psi(r)=B_{1} \frac{(i \mu / 2)^{l+\frac{1}{2}}}{\Gamma\left(l+\frac{3}{2}\right)} \cdot r^{l+1}-B_{2} \frac{\Gamma\left(l+\frac{1}{2}\right)}{\pi(i \mu / 2)^{l+\frac{1}{2}}} \cdot r^{-l}
\end{equation}
for the radial eigenfunction of the non-minimally coupled massive scalar field configurations in the intermediate radial region
\begin{equation}
{e}^{-4\alpha \Phi /(n-1)}\sqrt{\alpha} Q \ll r_{+}\ll 1 / (\frac{4}{n-1}y) 
\end{equation}
In both cases, A2 will develop to be much larger than A1 in the end, and there is a potential well outside the event horizon at this time, that is, superradiant instability will occur at that time.

\section{Summary and discussion}
In this article, we introduce $\mu=y\omega$\cite{20,21} into the  dilaton black hole, and discuss the superradiation instability of the  dilaton black hole.

  When y is greater than some limit, the effective potential of the equation exists at the poles, then a potential well appears outside the event horizon, when ${e}^{-4\alpha \Phi /(n-1)}\sqrt{\alpha} Q \ll r_{+}\ll 1 / (\frac{4}{n-1}y) $, so\ the charged dilaton black hole is superradiantly unstable at that time.

{\bf Acknowledgements:}\\
This work is partially supported by  National Natural Science Foundation of China(No. 11873025).

\end{document}